\documentclass[11pt,a4paper,twoside]{article}

\usepackage{amssymb}
\usepackage{amsmath}
\usepackage{bm}
\usepackage{graphicx}
\usepackage{subfigure}
\usepackage{hyperref}
\usepackage{color}
\usepackage{revsymb}
\usepackage[font=small,labelfont=bf]{caption}
\usepackage[title]{appendix}

\usepackage{booktabs}
\usepackage{array}
\usepackage{enumitem}
\usepackage{cite}
\usepackage[affil-it,auth-lg]{authblk}
\usepackage{hyperref}
\usepackage{eso-pic}

\newcolumntype{K}[1]{>{\centering\arraybackslash}p{#1}}

\begin{document}
	
	\title{Relativistic Gravitational Collapse by Thermal Mass}
	
	\author[1]{Zacharias Roupas\thanks{Zacharias.Roupas@bue.edu.eg}}
	\affil[1]{Centre for Theoretical Physics, The British University in Egypt,
		Sherouk City 11837, Cairo, Egypt
	} 
	
	\date{}
	
	\maketitle
	
	\begin{abstract}
		Gravity and thermal energy are universal phenomena which compete over the stabilization of astrophysical systems. The former induces an inward pressure driving collapse and the latter a stabilizing outward pressure generated by random motion and energy dispersion. Since a contracting self-gravitating system is heated up one may wonder why is gravitational collapse not halted in all cases at a sufficient high temperature establishing either a gravo-thermal equilibrium or explosion. Here, based on the equivalence between mass and energy, we show that there always exists a temperature threshold beyond which the gravitation of thermal energy overcomes its stabilizing pressure and the system collapses under the weight of its own heat. 
	\end{abstract}

	Mass and energy are equivalent and gravitate irrespective from their origins. Likewise, motion of material particles and transmission of energy by field waves are phenomena that manifest the universal nature of heat. Thus, gravity and thermal energy are the two universal properties of physical reality known as yet, while they are ultimately tightly connected.  Gravity causes motion and vice versa. 
	Nevertheless, they have counter primary effects on the stability of systems. Gravity drives contraction, while thermal pressure expansion. 
	
	When a system contracts adiabatically it is heated. 
	Since thermal energy gravitates there might exist a threshold upper temperature beyond which gravity of heat overcomes its outward pushing pressure. This is the question we address. The answer may be emphasized, by using heuristically only three basic equations of physics; Einstein's equivalence between mass and energy $m=E/c^2$, Newton's law of gravitational energy $E_{\rm gravity} \sim - GM^2/R$ and the equation of internal thermal energy $E_{\rm thermal } \sim NkT$. Assuming that thermal energy gravitates and since gravitational energy grows to the square of mass it follows that gravitational energy grows to the square of temperature. However, thermal energy grows only linearly with temperature. Thus, there should exist a temperature at which gravitational energy induced by thermal mass overcomes thermal energy. 
 Consider in particular a spherical self-gravitating system with radius $R$ and mass $M = Nm + E_{\rm thermal}/c^2$ consisting of the total rest mass of $N$ constituents with mass $m$ and the thermal mass $M_{\rm thermal} = E_{\rm thermal}/c^2$. The self-gravitational energy of the system to first order in $(GM/Rc^2)$\footnote{Weinberg, as in section 11.1 of Ref. \cite{Weinberg_1972gcpa.book} has shown that the relativistic self-gravitational energy $E_{\rm gravity}$ may be expanded in terms of $G\mathcal{M}(r)/rc^2$ as 
$
 E_{\rm gravity} = - \int_0^R 4\pi r^2 
 \left\lbrace 
 	\frac{G\mathcal{M}(r)}{rc^2} + \frac{3}{2}\left(\frac{G \mathcal{M}(r)}{r c^2}\right)^2 + \mathcal{O}\left((G\mathcal{M}/rc^2)^3\right)
 \right\rbrace \rho(r) c^2 dr,
 $
 where $\rho(r)$ denotes the mass density at radius $r$ and $\mathcal{M}(r)$ denotes the total mass-energy included in radius $r$. Since $\rho \sim M$, the relativistic self-gravitational energy of a static body is of order $E_{\rm gravity} \sim \mathcal{O}\left( M^2 \right)$.}
is proportional to
	\begin{equation}
		E_{\rm gravity} \sim -G\frac{M^2}{R} = -N^2\, G \frac{m^2}{R}\left(1 + \frac{kT}{mc^2} \right)^2.
	\end{equation}
	The gravitational energy overcomes the thermal pressure and the system gets destabilized if
	\begin{equation}\label{eq:ineq}
		\left| E_{\rm gravity}\right| > E_{\rm thermal} 
		\Rightarrow
		 \left(1 + \frac{kT}{mc^2} \right)^2
		 > \frac{Rc^2}{GM_{\rm rest}}\frac{kT}{m c^2}.
		\end{equation}
		The smaller root of this inequality, call it $T_1$, suggests that for sufficiently low temperature $T<T_1$ the gravitational energy overcomes thermal energy as expected (gravothermal catastrophe). The new element put forward here is that at sufficiently high temperature the gravitational energy is dominated by thermal mass. There exists therefore also an upper threshold, let us call it $T_{\rm c}$, steaming from the bigger root of (\ref{eq:ineq}) that is generated by the gravitation of thermal mass. 
		For any 
		\begin{equation}
			T > T_{\rm c}
		\end{equation}
the outward pointing pressure of thermal energy is overcome by its own gravitation and the system gets destabilized.
The characteristic temperature for $kT/ mc^2 \gg 1$ equals
\begin{equation}\label{eq:T_c}
			kT_{\rm c} \simeq mc^2
			\frac{Rc^2}{GM_{\rm rest}} .
		\end{equation}
	Therefore, by this simple analysis it is in addition predicted that this characteristic temperature depends on the compactness of rest mass $GM_{\rm rest}/Rc^2$. 
		
\begin{figure}[tb]
			\begin{center}
				\subfigure[Double spiral of the carolic curve]{ \label{fig:DoubleSpiral_TvsE}
				\includegraphics[scale = 0.5]{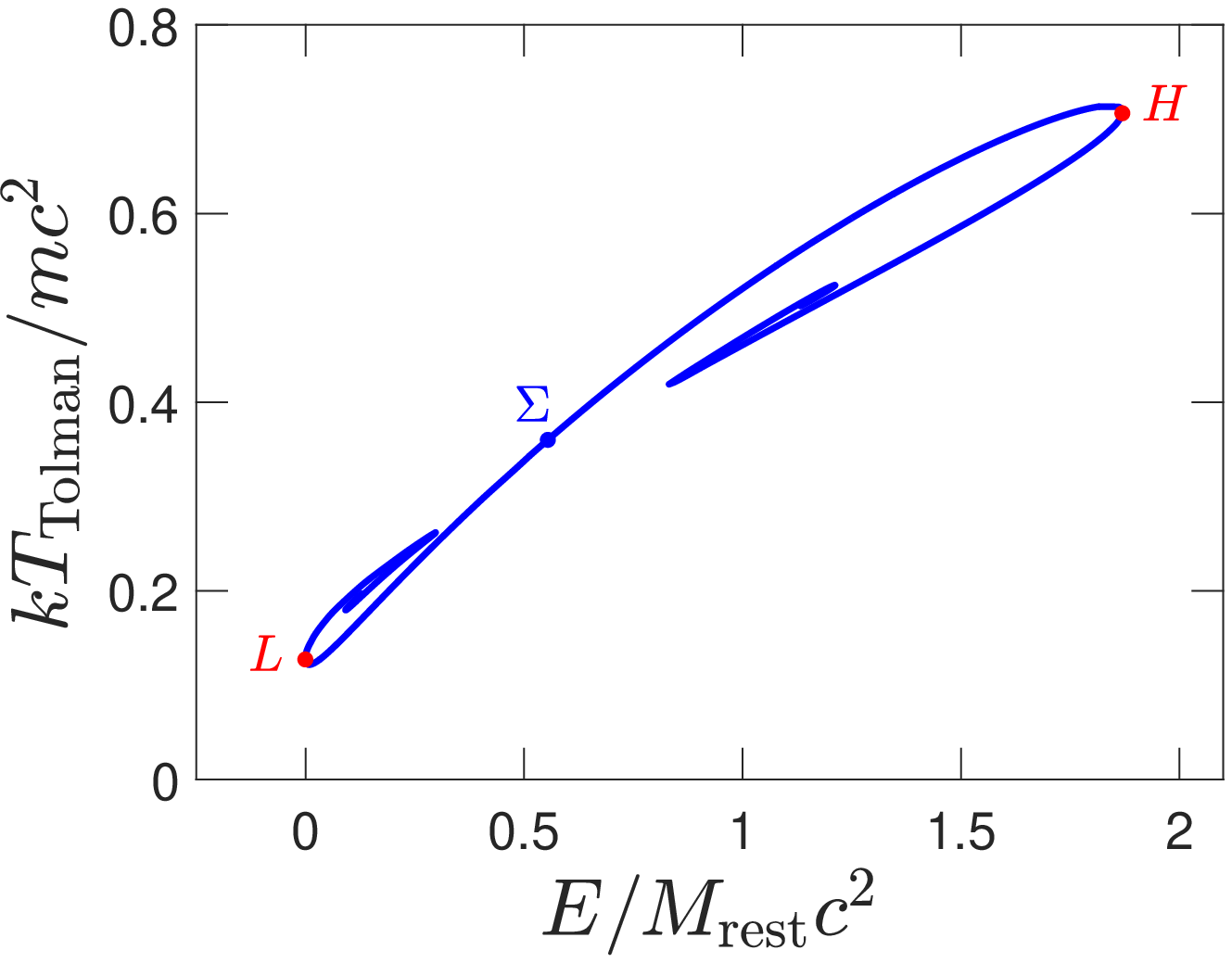} }
				\subfigure[Density contrast]{ \label{fig:density_contrast}
	\includegraphics[scale = 0.5]{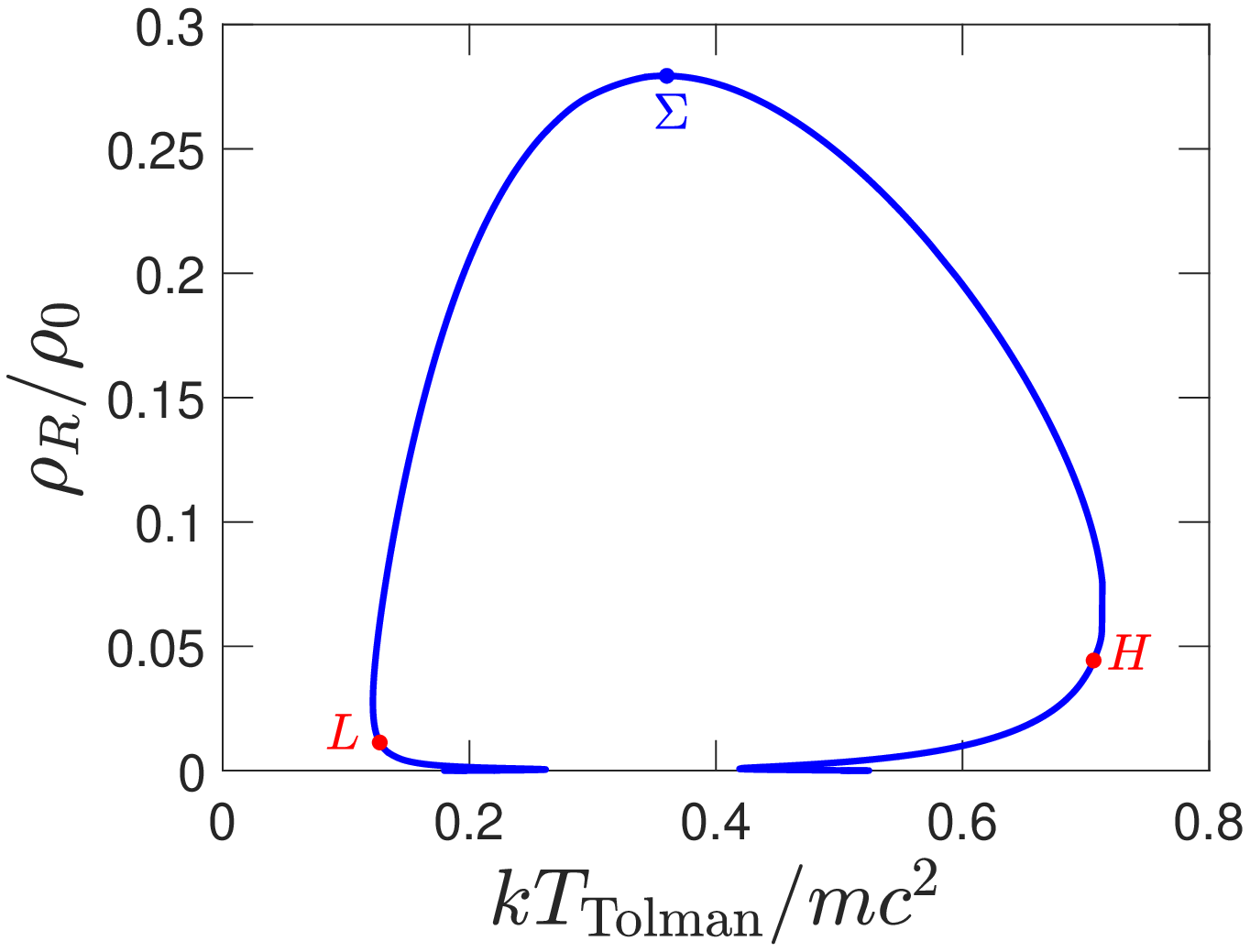} }
				\caption{ The series of hydrostatic, thermal equilibria of  the relativistic self-gravitating ideal gas for $GM_{\rm rest}/Rc^2 = 1/4$. The branch $LH$ is stable, while all points not in this branch are unstable and the empty regions outside points $L$, $H$ signify the absence of equilibria. At point $\Sigma$ the gravitation of thermal mass takes over its outward pushing thermal pressure and any further temperature increase results 
					in decreasing the edge to central density instead of homogenizing the system. Point $L$ designates the relativistic generalization of the Newtonian gravothermal catastrophe and point $H$ the relativistic gravothermal instability caused by thermal mass. 
				}
				\label{fig:caloric+thermal_mass}
			\end{center} 
		\end{figure}

		The result (\ref{eq:T_c}) is astonishingly close to the exact one, obtained by a precise relativistic treatment \cite{Roupas_CQG_RGI_2015,Roupas_2015PRD,Roupas_2013CQG,Roupas_2015CQG_32k9501R,Roupas_Symmetry_2019}. We consider the equation of state of the relativistic ideal gas 
		\begin{equation}\label{eq:clas_eos}
	P(\rho(r),T(r)) = \frac{\rho(r) c^2}{b(r)(1+\mathcal{F}(b(r)))},\quad \mathcal{F}(b) = \frac{K_1(b)}{K_2(b)} + \frac{3}{b} - 1 \; \text{and}\;
	b=\frac{mc^2}{kT}, 
		\end{equation}
		where $K_\nu(b)$ are the modified Bessel functions, $P$ is the pressure and $\rho$ is the total mass density. The equation of state may also be written as $P=(\rho_{\rm rest}/m)kT$ where $\rho_{\rm rest}$ is the rest mass density. Keeping the rest mass compactness fixed at several prescribed values, we solve the relativistic equation of hydrostatic equilibrium and Tolman equation, respectively,
		\begin{align}
		\label{eq:TOV}
		\frac{{\rm d}P(r)}{{\rm d}r} &= -\left(\rho (r) + \frac{P(r)}{c^2}\right)\left( \frac{G\mu(r)}{r^2} + 4\pi G \frac{P(r)}{c^2}r\right)
		\left( 1 - \frac{2G\mu(r)}{rc^2} \right)^{-1}, 
		\\
		\label{eq:Tprime}
		\frac{{\rm d}T(r)}{{\rm d}r} &= \frac{T(r)}{P(r) + \rho(r) c^2} \frac{{\rm d}P(r)}{{\rm d}r}, 
		\end{align}
		where $\mu(r)$ is the total mass-energy contained within radius $r$. Indeed, we find that the characteristic temperature depends on the rest mass compactness of the system as predicted by our previous heuristic analysis.
		In Figure \ref{fig:DoubleSpiral_TvsE} is plotted the caloric curve for $GM_{\rm rest}/Rc^2 = 1/4$, a typical rest mass compactness value of cool neutron stars. Each point corresponds to a thermodynamic equilibrium configuration. All points in the branch $LH$ are stable, while all other branches  (the two spirals) are unstable and the empty regions signify the complete absence of any kind of equilibrium. In y-axis is denoted the Tolman temperature $T_{\rm Tolman}$, which is the global quantity that remains constant in equilibrium \cite{Tolman:1930}, while local temperature $T(r)$ in Gravity attains a radial gradient (\ref{eq:Tprime}) according to the Tolman-Ehrenfest effect \cite{Tolman-Ehrenfest:1930}. 
	
	The caloric curve has the form of a double spiral, discovered firstly in \cite{Roupas_CQG_RGI_2015}. As we increase the rest mass, the spiral shrinks and becomes a point for $GM_{\rm rest}/Rc^2 = 0.35$. No equilibria may be attained above this value.
		
		The double spiral manifests two kinds of gravothermal instabilities. The low-energy one designated with point $L$ is the relativistic generalization of Antonov instability, called also `gravothermal catatrophe' \cite{antonov,lbw}. This instability is of Newtonian origin and is caused due to the decoupling of a self-gravitating core with negative specific heat capacity from a halo with positive specific heat capacity. A heat transfer from the core to the halo cannnot be reversed. This low-energy instability in adiabatic conditions occurs as the system gets expanded and not contracted. The subsequent cooling results in weakening heat's ability to sustain the gravity of rest mass.
		
		The upper spiral on the other hand designates a purely relativistic instability caused by the weight of heat during contraction and heating \cite{Roupas_CQG_RGI_2015,Roupas_Symmetry_2019}. In Figure \ref{fig:density_contrast} it is evident that starting from the equilibrium $L$ and heating up the system a little, the density contrast, namely the edge over the central density of the sphere, increases. The system becomes more homogeneous as expected because of the increase of thermal energy over gravitation. However, at a certain temperature, designated by point $\Sigma$, a peculiar reverse occurs. The system becomes less homogeneous while heated! The gravity of thermal energy dominates over its outward pointing pressure. Heating the system further results only in condensing the system and the appearance of the relativistic gravothermal instability, designated by point $H$, that is caused by the weight of heat.
		
		This result suggests that the gravitational collapse to a black hole cannot be halted by thermal energy alone. When all fuels and other stabilizing mechanisms are exhausted, thermal energy cannot come to the rescue no matter how hot the collapsing system becomes. 
		
		This phenomenon especially applies to supernova. Equation (\ref{eq:T_c}) and Figure \ref{fig:DoubleSpiral_TvsE} suggest that
		$kT_{\rm c}$ is expected to be of the same order with $mc^2$ for a supernova core, where $m$ should be identified with the rest mass of the primary gas constituent. If the remnant core attains a temperature and density that allows QCD deconfinent \cite{10.1111/j.1365-2966.2004.07849.x,PhysRevD.84.085003,2011A&A...528A..71B,Roark_PhysRevC.98.055805}, $m$ can be identified with the rest mass of up quark or even strange quark at higher densities \cite{Fraga_2014,Baym_2018} giving $kT_{\rm c} \sim 10-100 {\rm MeV}$. These values should designate the maximum possible temperature of a supernova core that may form a hot protoneutron star. Indeed, a protoneutron star is believed to attain temperatures $\sim 10MeV$ \cite{Burrows_2013RvMP...85..245B}. If the core attains higher temperature, it will continue to collapse under the weight of its own heat. Because of the core's negative specific heat capacity, even if significant amount of radiation is emitted from the core it will only result in further increasing its temperature and thus its gravitational thermal mass.
				 		
The relativistic gravothermal instability shall in addition have implications for cosmology. In an open Universe, an early region which becomes the observable Universe cannot be arbitrarily dense and hot in the past. For a certain compactness there always exists some temperature above which the system cannot expand because of the extra gravity of heat. The same is true for density fluctuations in an open Universe and for the whole patch of a closed Universe. Nevertheless, the expansion adds an additional inertial force, pushing outwards, which should be taken into account. 

We highlighted here some basic underlying physics of gravitational instability that has been, as yet, evading attention. This is that random motion and energy dispersion can generate such intense gravity that overcomes its own stabilizing, against gravitational collapse, pressure. This effect implies the inevitability of black hole formation and shall dominate its very late stages.

\section*{Acknowledgements}
The author thanks Amr El-Zant for discussions on cosmology.
	
\bibliography{Roupas_GRF_2020}

\bibliographystyle{unsrt}


\end{document}